\documentclass[conference]{IEEEtran}

\usepackage{amsmath}
\usepackage{amssymb}
\usepackage{booktabs}
\usepackage{graphicx}
\usepackage{url}
\usepackage[hidelinks]{hyperref}

\title{ToolMisuseBench: An Offline Deterministic Benchmark for Tool Misuse and Recovery in Agentic Systems}

\author{\IEEEauthorblockN{Akshey Sigdel}
\vspace{-12px}\\\IEEEauthorblockA{Independent Researcher}\vspace{-12px}\\aksheysigdel@u.boisestate.edu
\and
\IEEEauthorblockN{Rista Baral}
\vspace{-12px}\\\IEEEauthorblockA{Independent Researcher}\vspace{-12px}\\ristabaral@u.boisestate.edu}

\begin{document}

\maketitle

\begin{abstract}
Tool using agents often fail for operational reasons even when language understanding is strong. Common causes include invalid arguments, interface drift, weak recovery, and inefficient retry behavior. We introduce ToolMisuseBench, an offline deterministic benchmark for evaluating tool misuse and recovery under explicit step, call, and retry budgets. The benchmark covers CRUD, retrieval, file, and scheduling environments with replayable fault injection. It reports success, invalid call behavior, policy violations, recovery quality, and budgeted efficiency. We release a public dataset with 6800 tasks and a reproducible evaluation pipeline. Baseline results show fault specific recovery gains for schema aware methods, while overall success remains limited under the released authorization and hard failure settings.
\end{abstract}

\section{Introduction}
Tool integrated language agents are now used in workflows where action reliability is as important as textual quality. In these settings, failure often comes from malformed tool calls, interface mismatch, and unstable post failure behavior rather than pure reasoning error. A useful benchmark must therefore support reproducibility, clear failure attribution, and controllable stress conditions. ToolMisuseBench targets this reliability gap and complements broad capability suites such as HELM and BIG bench \cite{helm,bigbench}.

ToolMisuseBench uses deterministic simulators and declarative fault plans, so each episode is exactly replayable under a fixed seed. This enables direct comparison across agents and robust ablations. Faults are separated into schema drift, rate limit, timeout, authorization related failures, and adversarial error rewriting, which isolates failure origin at call construction, policy handling, or recovery.

Our contributions are threefold. First, we define a model agnostic benchmark specification with explicit budgets and structured success checks. Second, we provide a unified open pipeline for generation, evaluation, scoring, and artifact production. Third, we release a large public dataset and baseline results that establish a reproducible reference point. The key novelty is deterministic replayable fault injection with explicit recovery and budget metrics in a fully offline setting.

The motivation for this benchmark is practical as well as methodological. In real deployments, teams often observe a mismatch between perceived model intelligence and observed operational reliability. Agents that appear strong in static prompting can still fail repeatedly once strict schemas and budget limits are enforced. A benchmark that is insensitive to these conditions can overestimate readiness. We therefore emphasize measurable operational behavior and not only final answer quality.

A second motivation is comparability over time. Reliability work typically advances through many small changes in prompting, repair policy, and retry logic. Without replayable environments, it is difficult to attribute gains to specific changes. ToolMisuseBench is intended to support this iterative process. Because episodes are replayable, the same tasks can be reused for controlled longitudinal evaluation as methods evolve.

\section{Related Work}
Benchmark design for language agents has advanced rapidly, yet many suites prioritize broad capability over controlled tool reliability. General benchmarks remain essential \cite{helm,bigbench}, but they do not isolate schema errors, interface drift, and recovery under strict budgets. Agent suites increase interaction realism \cite{agentbench}, but deterministic replay of failure conditions is still limited.

Tool use methods improve completion and grounding when interfaces are stable \cite{toolformer,react,gorilla}. Recent controlled studies also show that schema based tool contracts can reduce interface misuse without fully resolving semantic failures under tight budgets \cite{schemafirst}. Deployment oriented studies report operational failures such as malformed calls, policy violations, and brittle retries \cite{apibank}. Complementary policy first work argues for explicit enforcement layers around tool orchestration rather than model only mitigation \cite{guardrailsinfra}. ToolMisuseBench complements these lines of work by centering deterministic failure injection and explicit recovery metrics for repeatable reliability analysis.

\section{Benchmark Formulation}
Each task is a structured record with instruction, tool schemas, initial state, success criteria, fault plan, and budget limits. Budgets include maximum steps, tool calls, retries, and timeout metadata. Evaluation runs as an episode where the agent repeatedly observes state and issues one tool call per step until completion or termination.

Fault controllability is central. Fault plans are per task and replayed deterministically. Supported faults are schema drift, rate limit, timeout, authorization related failures, and adversarial error rewriting. Schema drift mutates interface expectations, execution faults emulate service instability, and adversarial errors test robustness when feedback is ambiguous.

Success uses state based and transcript based checks. State checks include equality, existence, membership, and keyed dictionary value checks. Transcript checks enforce minimum counts for tool calls and successful tool calls. This supports both state transition and interaction completion tasks. A task is successful only when all criteria are satisfied.

This formulation provides two useful properties for analysis. The first property is decomposition. Researchers can inspect whether a failure originates from invalid call construction, policy rule violation, or inability to recover after receiving structured feedback. The second property is controlled hardness. By adjusting fault plans and budgets while keeping domain semantics fixed, one can increase stress without changing the underlying task family.

The benchmark intentionally avoids hidden reward shaping or opaque environment heuristics. Every success criterion and every fault trigger is represented in explicit task metadata. This makes each evaluation episode auditable and makes disagreement resolution straightforward when comparing methods across research groups.

The benchmark is budget aware by construction. Let $N$ denote the number of evaluated tasks and let $k$ denote a tool call cap. Budgeted success for an agent is computed as
\begin{equation}
S(k) = \frac{1}{N}\sum_{i=1}^{N}\mathbb{1}\left[\mathrm{TaskSuccess}_i=1 \land \mathrm{ToolCallsUsed}_i\leq k\right].
\end{equation}
The area under this curve summarizes efficiency aware performance and prevents unlimited retry behavior from masking instability.

\section{System Implementation}
ToolMisuseBench is implemented as a Python package with a single command line workflow for generation, evaluation, and experiment reproduction. Core environments are deterministic simulators for CRUD, retrieval, file, and scheduling tasks. Each environment enforces schema validation and applies the fault engine before execution.

The fault engine applies trigger logic and fault behavior in a fixed order. Triggers include tool targeting, nth call targeting, argument pattern targeting, and seeded probabilistic activation. Error handling includes retry accounting, termination on retry overflow, and structured traces. Replayability is guaranteed through task seeded random streams and declarative trigger definitions.

The same codebase supports built in baselines and external custom agents. Custom agents are loaded by module path and evaluated through the same harness and scoring path. The external interface is minimal and requires only reset and act methods, which lowers integration cost while preserving protocol consistency.

From an engineering perspective, the implementation is designed for reproducibility first and convenience second. All key artifacts are generated by stable commands, including task files, trace files, aggregate reports, table files, and figure data files. This reduces divergence between internal analysis notebooks and published results. It also makes it easier for independent users to verify claims without reimplementing evaluation logic.

The implementation also treats error transparency as a first class requirement. Structured error payloads preserve fault context for each failed step, including trigger information and transformed feedback under adversarial settings. This enables fine grained post hoc analysis of repair behavior and helps identify whether an apparent improvement is robust or merely exploitative of a narrow feedback pattern.

\section{Dataset Construction and Quality Assurance}
The released dataset contains 6800 tasks with split sizes of 5000 for train, 800 for development, and 1000 for public test. Domains are balanced across CRUD, retrieval, files, and scheduling. The generator supports default sizing, profile based scaling, and explicit split overrides.

Per task coherence is enforced during generation through validation of instruction presence, domain state consistency, criteria structure, and fault references against available tools. Generation fails on any invalid record. Artifacts include a checksum manifest and version freeze metadata.

The large split shows strong variation in instruction text, initial states, and success criteria while preserving deterministic replay. A quality report utility summarizes uniqueness, domain balance, fault balance, and duplicate identifier checks. On the released split, train instruction uniqueness is 0.5622, train initial state uniqueness is 0.7510, and duplicate task identifiers are zero across all splits.

Data quality decisions were made to balance diversity and control. We introduce substantial lexical and structural variation in instruction text and state initialization, while keeping tool semantics stable enough for meaningful cross agent comparison. This avoids both extremes where tasks become nearly identical or excessively noisy. As a result, the dataset can support robust aggregate reporting while still exposing many distinct failure contexts.

The split design also reflects practical use. The train split is large enough to support method development and debugging at scale, while the development and public test splits remain tractable for repeated benchmarking. Because all splits are produced by the same deterministic generator family, users can create additional internal splits with consistent semantics when needed.

\section{Experimental Protocol}
Evaluation is performed on the public test split with three baselines. The public test split contains 1000 tasks under the released large profile. The first baseline is a deterministic heuristic policy. The second adds schema based repair for common argument errors. The third adds a lightweight policy awareness layer on top of schema repair. All baselines run under identical budgets, environments, and fault settings.

Reported metrics include task success, policy violations, invalid call rate, recovery success, time to recovery, tool calls used, budget exceeded rate, and catastrophic failure rate. We also compute budgeted success at caps 4, 8, 16, and 32 with normalized area under curve. Each run outputs aggregate summaries and per task traces for direct failure analysis.

The protocol is intentionally strict about comparability. No baseline receives privileged environment access, and all baselines consume the same observation interface. Agent side variation is limited to decision logic. This ensures that measured differences reflect behavioral policy changes rather than hidden instrumentation advantages.

In addition to aggregate scores, the protocol encourages fault conditioned analysis. Per fault views reveal patterns that can disappear in aggregate averages. This is important for reliability work, because small global improvements can hide severe regressions on specific failure classes.

\section{Results}
Table~\ref{tab_overall} summarizes aggregate baseline behavior. Overall task success is 0.25 for all three baselines on the current test setting. The heuristic baseline uses fewer tool calls on average, while schema aware and policy aware baselines show identical aggregate scores in this release. Recovery success increases for schema aware and policy aware baselines in timeout affected tasks, but global success remains bounded by the released fault mixture and budget pressure.

Per fault analysis shows a clear pattern. Timeout tasks are more recoverable than the released rate limit and authorization settings under current baselines. Schema drift tasks show partial recovery with elevated policy violations in repair based agents, which suggests over correction under severe interface drift. Budgeted success curves are flat with AUC 0.25 for all agents, indicating that policy and planning quality, not cap scaling alone, is the dominant bottleneck.

Table~\ref{tab_fault} highlights fault specific behavior for success and recovery. The heuristic baseline has zero recovery on timeout and schema drift settings despite moderate success on selected tasks. Schema repair and policy aware baselines reach 0.5015 recovery on timeout and 0.4970 on schema drift. Policy violations also increase on schema drift for repair based methods, which motivates selective repair strategies with stronger safety constraints. By contrast, all baselines remain at zero measured success on the released rate limit and authorization subsets in this configuration.

These results suggest that recovery heuristics are currently effective only in bounded contexts where failure feedback remains tractable. Under the released configuration, once failures involve authorization constraints or persistent rate limiting, simple repair loops provide limited benefit. The identical schema repair and policy aware scores in this release also indicate that the added policy layer is too narrow to change aggregate behavior under the current task mix. This distinction is important for future method design. Stronger systems likely require policy aware replanning that can switch goals or select lower risk fallback actions under budget pressure.

A second observation is that call efficiency alone does not predict reliability. The heuristic baseline uses fewer calls but does not improve overall task success. Repair based methods use more calls and improve recovery for selected faults, yet they do not improve global success under the current fault mixture. This supports the view that reliable tool use is a joint optimization problem over correctness, safety, and budget discipline.

\begin{table}[t]
\centering
\small
\caption{Overall metrics on the public test split}
\label{tab_overall}
\begin{tabular}{@{}lcccc@{}}
\toprule
Agent & Success & Violations & Recovery & Calls \\
\midrule
Heuristic & 0.250 & 0.168 & 0.000 & 2.95 \\
Schema repair & 0.250 & 0.166 & 0.250 & 3.25 \\
Policy aware & 0.250 & 0.166 & 0.250 & 3.25 \\
\bottomrule
\end{tabular}
\end{table}

\begin{table}[t]
\centering
\small
\caption{Fault specific success and recovery rates}
\label{tab_fault}
\begin{tabular}{@{}lccc@{}}
\toprule
Setting & Heuristic & Schema repair & Policy aware \\
\midrule
Timeout success & 0.499 & 0.502 & 0.502 \\
Timeout recovery & 0.000 & 0.502 & 0.502 \\
Schema drift success & 0.503 & 0.497 & 0.497 \\
Schema drift recovery & 0.000 & 0.497 & 0.497 \\
Authz success & 0.000 & 0.000 & 0.000 \\
Rate limit success & 0.000 & 0.000 & 0.000 \\
\bottomrule
\end{tabular}
\end{table}

\section{Discussion and Limitations}
The benchmark enables deterministic failure attribution and fair side by side evaluation. Current baseline results show fault specific gains and also show that recovery alone is insufficient when the released policy related and hard failure settings dominate. This outcome motivates integrated planning with selective fallback and aligns with recent policy first evidence that explicit enforcement infrastructure can materially change safety utility tradeoffs in tool orchestrated workflows \cite{guardrailsinfra}.

The benchmark is synthetic and intentionally controlled. This improves comparability but does not cover the full semantic range of production tools. Future versions should include richer mixed domain workflows, longer horizon dependencies, and broader policy rules. Baseline diversity is also limited and should be expanded with stronger learned agents and model adapters. In particular, the current policy aware baseline should be interpreted as a lightweight safety heuristic rather than a comprehensive policy reasoning system. Simulator based evidence should be treated as complementary to live system evaluation.

Another limitation concerns external cost realism. Although the benchmark models budgets and retries, it does not currently model monetary pricing, network jitter distributions, or heterogeneous service level guarantees. These factors matter in production. Future releases can incorporate richer budget semantics and calibrated latency models while preserving deterministic replay.

We also note that failure taxonomies evolve with tooling ecosystems. New interface patterns and safety policies may introduce additional failure modes beyond the current five fault families. Recent evidence on schema first tool design suggests that interface formalization and validation diagnostics deserve independent treatment alongside execution faults and recovery failures \cite{schemafirst}. The benchmark is designed to accommodate extension, and future work should evaluate expanded fault taxonomies to preserve relevance across deployment domains.

\section{Reproducibility and Artifacts}
We release the full benchmark implementation, generator, evaluator, and experiment pipeline as open source code.\footnote{Code repository: \url{https://github.com/akgitrepos/toolmisusebench}.} The public dataset release is available through Hugging Face \cite{hfdata}. The dataset includes train, development, and public test splits with checksum manifest and version freeze metadata. Reproduction requires only local execution with fixed seeds and no external runtime services.

The workflow is intentionally simple. Users generate data, evaluate built in or custom agents, and produce tables and figure data through a small set of commands. This supports transparent verification and lowers friction for third party comparison. The same process is used for our reported results. The numeric values reported in Tables~\ref{tab_overall} and \ref{tab_fault} are produced directly from the released experiment artifacts generated by this pipeline.

\section{Conclusion}
ToolMisuseBench provides a reproducible benchmark for studying tool misuse and recovery in agentic systems. Deterministic simulators, declarative fault plans, and explicit budget metrics enable controlled robustness analysis that is difficult to obtain from ad hoc evaluations. We release a large dataset, an end to end evaluation harness, and baseline results in one open pipeline. We expect this benchmark to support future work on reliable tool integration, safer repair policies, and stronger policy aligned planning under realistic failure conditions.

\end{document}